\documentclass[twocolumn]{aastex63}

\usepackage[T1]{fontenc}
\usepackage[utf8]{inputenc}

\graphicspath{{./figures/}}

\newcommand{\citeg}[1]{\citep[e.g.,][]{#1}}
\newcommand{\e}[1]{\times 10^{#1}}

\newcommand{\rev}[1]{#1} 
\newcommand{\revdel}[1]{} 

\usepackage{url}
\usepackage[english]{babel}
\usepackage{amsmath}
\usepackage{amsfonts}
\usepackage{hhline}
\hypersetup{unicode=true, pdftoolbar=true, pdfmenubar=true, pdffitwindow=false, pdfstartview={FitH}, pdftitle={Machine learning inference of the interior structure of low-mass exoplanets}, pdfauthor={Philipp Baumeister, Sebastiano Padovan, Nicola Tosi, Gr\'egoire Montavon, Nadine Nettelmann, Jasmine MacKenzie, Mareike Godolt},	pdfsubject={},	pdfcreator={\LaTeX\ with package \flqq hyperref\frqq}, pdfproducer={pdfTeX \the\pdftexversion.\pdftexrevision}, pdfkeywords={machine learning, exoplanets, neural networks, planet interior}, pdfnewwindow=true, colorlinks=true,linkcolor=.,citecolor=.,filecolor=magenta,urlcolor=blue}

\usepackage[capitalise]{cleveref}
\Crefname{figure}{Figure}{Figures}
\Crefname{section}{Section}{Sections}
\Crefname{subsection}{Section}{Sections}
\Crefname{subsubsection}{Section}{Sections}

\received{August 30, 2019}
\revised{November 15, 2019}
\accepted{November 27, 2019}
\submitjournal{ApJ}

\shorttitle{Machine learning inference of exoplanet interiors}
\shortauthors{P. Baumeister et al.}

\begin{document}

\title{Machine learning inference of the interior structure of low-mass exoplanets}

\correspondingauthor{Philipp Baumeister}
\email{philipp.baumeister@tu-berlin.de}

\author[0000-0001-9284-0143]{Philipp Baumeister}
\affiliation{\rev{Centre} of Astronomy and Astrophysics, Technische Universität Berlin, Hardenbergstraße 36, D-10623 Berlin, Germany}
\author[0000-0002-8652-3704]{Sebastiano Padovan}
\affiliation{Institute of Planetary Research, German Aerospace Center (DLR), Rutherfordstraße 2, D-12489 Berlin, Germany}
\author[0000-0002-4912-2848]{Nicola Tosi}
\affiliation{\rev{Centre} of Astronomy and Astrophysics, Technische Universität Berlin, Hardenbergstraße 36, D-10623 Berlin, Germany}
\affiliation{Institute of Planetary Research, German Aerospace Center (DLR), Rutherfordstraße 2, D-12489 Berlin, Germany}
\author{Gr\'egoire Montavon}
\affiliation{Institute of Software Engineering and Theoretical Computer Science, Technische Universität Berlin, Marchstr. 23, D-10587 Berlin, Germany}
\author[0000-0002-1608-7185]{Nadine Nettelmann}
\affiliation{Institute of Planetary Research, German Aerospace Center (DLR), Rutherfordstraße 2, D-12489 Berlin, Germany}
\author{Jasmine MacKenzie}
\affiliation{\rev{Centre} of Astronomy and Astrophysics, Technische Universität Berlin, Hardenbergstraße 36, D-10623 Berlin, Germany}
\author[0000-0003-4770-8551]{Mareike Godolt}
\affiliation{\rev{Centre} of Astronomy and Astrophysics, Technische Universität Berlin, Hardenbergstraße 36, D-10623 Berlin, Germany}

\begin{abstract}
    We explore the application of machine learning based \rev{on} mixture density neural networks (MDNs) to the interior characterization of low-mass exoplanets up to 25 Earth masses constrained by mass, radius, and fluid Love number $k_2$. We create a dataset of 900\:000 synthetic planets, consisting of an iron-rich core, a silicate mantle, a high-pressure ice shell, and a gaseous H/He envelope, to train \rev{a} MDN using planetary mass and radius as inputs to the network. \rev{For this layered structure,} we show that the MDN is able to \rev{infer the distribution of possible  thicknesses of each planetary layer from mass and radius of the planet. This approach obviates the time-consuming task of calculating such distributions with a dedicated set of forward models for each individual planet}. \revdel{This requires only the trained network, without the need to create a dedicated set of models for each given object to compare with the observations.}\rev{While gas-rich planets may be characterized by compositional gradients rather than distinct layers, the method presented here can be easily extended to any interior structure model.} The fluid Love number $k_2$ bears constraints on the mass distribution in the planets' interior and will be measured for an increasing number of exoplanets in the future. Adding $k_2$ as an input to the MDN significantly decreases the degeneracy of the possible interior structures. \rev{In an open repository we provide the trained MDN to be used through a Python Notebook.}
\end{abstract}


\section{Introduction}
	The characterization of the interior of observed exoplanets is one of the main goals in current exoplanetary science. With the large number of newly discovered exoplanets expected in the next ten years by ground-based surveys such as WASP \citep{pollacco2006WASPProject}, NGTS \citep{wheatley2017NextGeneration}, and HATNet/HATSouth \citep{hartman2004HATNETVariability, bakos2013HATSouthGlobal} as well as the ongoing TESS  space survey \citep{ricker2014TransitingExoplanet} and the upcoming PLATO mission \citep{rauer2014PLATOMission}, a rapid characterization scheme of the interior structure of these planets will become increasingly necessary to further our understanding of planetary populations. The vast majority of the confirmed exoplanets has been identified
    either through transits or radial velocities surveys.
    Planets identified with both techniques 
    are characterized by their mass and radius, which, combined, provide a first indication of the bulk composition through comparison with 
    theoretical mass-radius curves 
    \citep[e.g.,][]{valencia2006InternalStructure,sotin2007MassradiusCurve, zeng2013DetailedModel}. A common approach to the interior characterization of exoplanets is the use of numerical models to compute interior structures which comply with the measured mass and radius of the planet \citep[e.g.,][]{sotin2007MassradiusCurve, valencia2007DetailedModels, fortney2007PlanetaryRadii, wagner2011InteriorStructure, zeng2013DetailedModel, unterborn2019PressureTemperature}. As this is an inverse problem, it requires the calculation of a large number of interior models to obtain an overview over possible interior structures \rev{\citep{rogers2010FrameworkQuantifying, dorn2017GeneralizedBayesian, brugger2017ConstraintsSuperEarth}}.
    \rev{If other observables are used in addition to mass and radius, the number of samples needed for an accurate inference of possible interior structures increases drastically, due to the increase in dimensionality \citep[e.g.,][]{james2013IntroductionStatistical}. Thus, the inference can quickly become computationally expensive.}\revdel{Thus, sampling the solution space can be computationally expensive, especially if in addition to mass and radius other observables are used, which increases the volume of space to be sampled drastically \citep[e.g.,][]{james2013IntroductionStatistical}.} Moreover, with only mass and radius, possible solutions tend to be highly degenerate, with multiple, qualitatively different interior compositions that can match the observations equally well \citep[e.g.][]{rogers2010FrameworkQuantifying,rogers2010ThreePossible}.

    In this work, we explore a new approach to the interior characterization of exoplanets by employing a deep learning neural network to treat this inverse problem. In recent years, \revdel{Deep Neural Networks}\rev{deep neural networks} have been used in a number of exoplanetary science studies, with applications ranging from the detections of planetary transits \citep{pearson2018SearchingExoplanets, shallue2018IdentifyingExoplanets, chaushev2019ClassifyingExoplanet} to atmospheric 
	composition retrieval from measured planet spectra \citep{zingales2018ExoGANRetrieving, marquez-neila2018SupervisedMachine}, and the computation of critical core and envelope masses of forming planets \citep{alibert2019UsingDeep}. 
    We approach the characterization of exoplanets by first creating a large dataset of synthetic planets and then training a multitask \citep{caruana1997MultitaskLearning} mixture density network (MDN, \citealp{bishop1994MixtureDensity}) to infer plausible thicknesses of the planetary layers using observables like mass and radius. Mixture density networks are a special case of neural networks, which predict the parameters of a Gaussian mixture distribution instead of single output values\rev{, enabling us to approximate the entire posterior probability density function at once}. This property allows them to work with inverse problems that conventional neural networks often cannot address well. \revdel{There are two main advantages in using \rev{Deep Learning} for the interior characterization of planets\rev{ compared to conventional methods, such as Markov Chain Monte Carlo}}\rev{Neural networks are an optimal choice to approach the interior characterization of planets}: First, as neural networks can essentially approximate any arbitrary non-linear relations between some input and output values \citep{cybenko1989ApproximationSuperpositions, goodfellow2017DeepLearning, lu2017ExpressivePower}, they are very well suited to model the relation between observables and interior structure for an arbitrary number of observable parameters, as the mapping can be difficult to quantify\revdel{ with conventional methods}, especially for a large number of model parameters \citep{bishop1995NeuralNetworks}. Secondly, \rev{the computation time needed by a well-trained neural network to make a prediction is in the order of milliseconds. Compared to conventional methods, such as Markov Chain Monte Carlo, this} \revdel{the use of a well-trained neural network} significantly reduces the computational time for finding valid interior structures, allowing for rapid classification of exoplanets. Additionally, with the fast training times of neural networks \rev{(less than 30 minutes with our setup, see end of Section \ref{sec:ML})}, the effect of different combinations of observables can be tested rapidly\rev{ (although validating the trained MDNs and optimizing the network architecture may still take a considerable amount of time)}.
    
    Beyond mass and radius, additional observables that may be informative for the interior structure include elemental abundances of the host star\rev{, which may be representative of those of the planet \citeg{dorn2015CanWe, dorn2017GeneralizedBayesian, brugger2017ConstraintsSuperEarth}}, atmospheric composition \citep[e.g.,][]{Madhusudhan2016}, and the planetary fluid Love numbers \citep{padovan2018MatrixpropagatorApproach}. Of these, the fluid Love numbers represent the most direct constraint \rev{on the interior structure}, given that they only depend on \rev{the radial density distribution in the interior} \revdel{concentration in the interior}\citep{kramm2011DegeneracyTidal}.
   
    \rev{Fluid Love numbers are a subset of the more 
    general set of Love numbers, which are 
    parameters describing how a planet responds to
    a perturbation (e.g., tidal and rotational
    deformations).
    They depend on the timescale of the perturbation and
    on the rheological properties of the interior (i.e.,
    radial density profile, elastic properties, 
    viscosity). 
    For example, the rotational figure of the Earth
    is relaxed, i.e., it attained hydrostatic equilibrium
    and accordingly, its shape can be parameterized using
    fluid Love numbers 
    \citep{padovan2018MatrixpropagatorApproach}.
    At the same time, on the shorter tidal lunar 
    and solar timescales, the response is described
    by tidal Love numbers \citep[e.g.,][]{Petit2010}.}

    The deformed shape of a transiting extrasolar planet will modify its transit light-curve, introducing a second-order effect which can in principle be observed \rev{\citep{akinsanmi_detectability_2019,Hellard2019}}. 
    Similarly, for some particular configurations \rev{\citep[e.g., a close-in planet with a distant companion,][]{mardling2007LongtermTidal}}, \rev{absidal} precession is mainly driven by the tidal interaction of the planet \rev{with its host star} \citep{Batygin2009}, and a value of the Love number of the planet can be inferred \citep{csizmadia2019EstimateLove}.  
    Thus, radial velocity measurements as well as transit observations can be used, under certain circumstances, to infer the value of the Love number $k_2$, a task that is currently possible only for close-in gas giants.
    The ever-increasing temporal baseline of exoplanets observations and sensitivity of ground-based and space telescopes will make the measurement of the Love numbers of extrasolar planets increasingly possible.
    \rev{For example, using the new Echelle SPectrograph 
    for Rocky Exoplanets and Stable Spectroscopic 
    Observations (ESPRESSO) mounted on the Very Large 
    Telescope facility of the European Souther Observatory 
    \citep{Pepe2010}, on a 10-year-long observing baseline, 
    one may infer the $k_2$ of a body with a mass 
    of 3 $M_{\oplus}$ (Sz. Csizmadia, 
    personal communication).
    The advent of larger telescopes such as the 
    Thirty Meter Telescope \citep[e.g.,][]{Skidmore2015}
    will further increase the sensitivity for smaller planets.}
    \rev{In the following, we assume that measurements of the fluid Love numbers become available in the future.}
    
\section{Methods}
    \subsection{Interior structure model}\label{sec:interior_model}
    We use a newly developed interior structure model named TATOOINE (Tool for ATmospheres, Outgassing and Optimal INteriors of Exoplanets)  to construct the planets used for the training data.
    A modelled planet consists of four compositionally distinct layers: An iron-rich core, a silicate mantle, an ice shell, and a H/He envelope. Each layer is characterized by its mass fraction, with the sum of the mass of all layers constrained to add up to the planet's total mass $M_p$. 
    
    The model calculates radial profiles of mass ($m$), density ($\rho$), and pressure ($P$) by solving the following three coupled equations: 
    \begin{equation}
        \frac{dm(r)}{dr} = 4\pi r^2 \rho(r), \label{eq:mass}
    \end{equation}
    %
    \begin{equation}
        \frac{dP(r)}{dr} = -\frac{G m(r) \rho(r)}{r^2},\label{eq:hydrostatic}
    \end{equation}
    \begin{equation}
        P(r) = f(\rho(r), T(r), c(r)). \label{eq:eos}
    \end{equation}
    \cref{eq:mass} expresses the mass \revdel{distribution} of a spherical shell\rev{ of thickness $dr$}; \cref{eq:hydrostatic} is the condition of hydrostatic equilibrium, where $G$ is the gravitational constant;  \cref{eq:eos} is an equation of state (EoS) where $f$ indicates a function that is specific to the material of each model layer and depends on the density, temperature $T(r)$, and composition $c(r)$ profiles.

    The model setup and parameters closely follow the modelling approach of \citet{sotin2007MassradiusCurve}, with the addition of a zero-temperature H/He envelope to account for gas-rich planets \citep{salpeter1967TheoreticalHighPressure, seager2007MassRadiusRelationships}. \rev{Using a zero-temperature EoS implies that the envelope thickness inferred from mass and radius of the planet will tend to be too large (Section \ref{sec:envelope}).}
    Below the H/He envelope, temperatures in the planet follow an adiabatic profile \rev{starting with a temperature of 300~K at the envelope-ice interface} in all models. 
    We further assume that all planets share the same Earth-like mineral composition of the mantle and core. \rev{The interior structure of gas-rich planets could be better described by a compositional gradient rather than by chemically-distinct layers, an approach that is best suited for predominantly rocky bodies. Nevertheless, the machine-learning based inference approach that we present below is general and can be easily adapted to any interior structure model (Section \ref{sec:ML}).} We focus on planets below 25 Earth-masses, as these show the greatest diversity in bulk density among the discovered exoplanets \citep[e.g.,][]{lopez2014UnderstandingMassRadius}, hinting at a wide range of interior structures.
    
    \subsubsection{Iron-rich core}
    
    Density profiles derived from seismic data show that the density of the Earth's core is about 5--10\% smaller than the density of pure hcp-iron, suggesting the presence of lighter elements 
    such as sulfur, oxygen, silicon, carbon, and hydrogen \citep[e.g.][]{Rubie2015}. We chose an equation of state for the core where iron is alloyed with sulphur, which is depleted in the Earth's mantle compared to solar abundances \citep{murthy1970ChemicalComposition, poirier1994LightElements}. Several experimental studies suggest that the melting temperature of the Fe-FeS alloy is low enough to allow the Earth's outer core to be molten under the probable thermodynamic conditions \citep[e.g.,][]{yoo1993ShockTemperatures,fei1997HighPressureIronSulfur,morard2008SituDetermination}. We assume our model planet to have a liquid core  composed of a Fe-FeS alloy with a molar fraction of 13\% FeS \citep{poirier1994LightElements, sotin2007MassradiusCurve}. For planets more massive than Earth, \citet{stixrude2014MeltingSuperearths} suggests that the iron cores of planets in the entire Super-Earth mass range (up to 10 Earth masses) are at least partially liquid, and mantle convection is vigourous enough to sustain a dynamo.
    We neglect an inner solid core, as the total core radius is not expected to be affected significantly since the density difference between the solid and liquid phase is small \citep{sotin2007MassradiusCurve}. The temperature- and pressure-dependent density is calculated using a Mie-Grüneisen-Debye EoS \citep[e.g.,][]{sotin2007MassradiusCurve, fei2016ThermalEquation}.
    
    \subsubsection{Silicate mantle} 
    \label{sec:silicate_mantle}
    The silicate mantle in our model is divided into an upper and a lower mantle, separated by a phase transition at 23~GPa. The lower mantle is composed of bridgmanite and magnesiowüstite. The upper mantle is composed of olivine and ortho-pyroxene enstatite. The ratio of mantle components is assumed to be Earth-like, with a molar Mg/Si ratio of 1.131 \citep{sotin2007MassradiusCurve}. Additionally, all mantle silicates \rev{form} a solid\rev{-state} solution of their respective Mg and Fe endmembers with the ratio determined by the Mg number (Mg\#), which is defined as the mole fraction Mg/(Mg+Fe) of the mantle silicates.
    We use an Earth-like value of 0.9 for the Mg\# in all modeled planets. Density profiles are calculated using a temperature-dependent 3rd order Birch-Murnaghan EoS.
    
    \subsubsection{High pressure ice layer}
    \rev{Many exoplanets have intermediate bulk densities, which may be indicative of large amounts of water in the interior} 
    \citep[e.g.,][]{leger2004NewFamily,adams2008OceanPlanet,Barr2018,Zeng2019}. We allow for the
    presence of water by using a layer of ice VII,
    which is the most stable water ice high-pressure phase over a wide range of pressures \citep{hemley1987StaticCompression}. 
    We neglect other water phases, as the density differences between solid ice phases are small \citep{hemley1987StaticCompression} and do not have a major effect on the thickness of this layer. By doing this, we neglect liquid and super-ionic phase effects, which would lead to more degenerate solutions of interior structures \citep{nettelmann2011ThermalEvolution}. Also, it must be noted that higher temperatures in the ice shell may significantly change the thickness of the layer \citep{thomas2016HotWater}.
    
    \subsubsection{H/He gas envelope}\label{sec:envelope}
    
    Many massive (> 10 M$_\oplus$) exoplanets probably have an extended primordial envelope consisting of hydrogen and helium \citep[e.g.,][]{lopez2014UnderstandingMassRadius}.
    To model this outer gaseous layer, we assume an envelope of solar-like composition with 71\% hydrogen and 29\% helium (by mass), and use a zero-temperature Thomas-Fermi-Dirac EoS from \citet{salpeter1967TheoreticalHighPressure} to model the pressure-dependent density. Using a zero-temperature envelope overestimates the density of the gas mixture\rev{ compared to more sophisticated temperature-dependent EoS. As a consequence, more of the planet needs to be taken up by the envelope to account for the planet's bulk density when modeling the interior.}
    \rev{Especially for planet models with extended atmospheres, this means that, when inferring the thickness of the envelope from mass and radius of the planet, the distribution of possible solutions is shifted towards higher envelope thicknesses.}\revdel{, thus establishing a lower bound on the envelope thickness \citep{seager2007MassRadiusRelationships}}\rev{ We define the planet's transit radius at a pressure of 100~mbar.}

    \subsection{Mixture density networks}\label{sec:ML}
    \rev{Feedforward neural networks are a fundamental part of machine learning algorithms. Their goal is to learn the mapping between inputs $x$ and corresponding outputs $y$ \citep[see  e.g.,][]{bishop2006PatternRecognition, goodfellow2017DeepLearning}. 
    A neural network consists of several layers of neurons. For a conventional feedforward neural network, each layer receives the outputs from the previous layer as input and computes a new set of outputs. This calculation is done by the neurons in the layer, where each neuron receives input from other neurons of the previous layer and computes an output value, using the so-called activation function. These functions are typically nonlinear, and thus enable the neural network to approximate any arbitrary, non-linear function. The first layer in a network receives data features as input and is therefore called the input layer, and the final layer is called the output layer because it outputs the predictions of the network. All layers between input and output layer are referred to as `hidden'. Feedforward networks are so named because information flows only in one direction from the input layer through the hidden layers to the output layer, without loops or feedback.}
    
    \rev{A neural network is trained with a set of training data that provides examples of which output values correspond to each set of input values. During training, these data are passed through the network, where the weighting of each neuron is iteratively adjusted so that the predicted value approaches the actual target value. This difference is measured by the so-called loss function. The goal of training is to minimize the loss function. Therefore, the gradient of the loss function at the current iteration is calculated by backpropagating the predicted values through the network \citep{bishop2006PatternRecognition, goodfellow2017DeepLearning}, and then using a gradient descent algorithm to optimize the model weights.}
    
    The conventional approach in neural network training is to minimize the mean squared error\rev{ between the known values from the training data and the values predicted by the network}, which results in these models approximating the \rev{average of the training data, conditioned on the input parameters}. For forward \rev{problems}, \rev{where a set of input parameters maps to a single set of output values, }this approach usually yields the desired solution. For inverse problems, where there may be multiple \rev{sets of output values for each feature}, these models are inadequate in many cases \citep{bishop1994MixtureDensity}. Mixture density networks (MDN) overcome this problem by combining conventional neural networks with a probability mixture model. While a conventional neural network maps inputs to deterministic output variables, \rev{a} MDN predicts the \rev{full conditional probability distribution for each input}. We use a Gaussian mixture model for our MDN, where the total distribution is the normalized linear combination of weighted normal distributions, as Gaussian mixtures can represent a wide range of data distributions \citep{bishop1994MixtureDensity, bishop2006PatternRecognition}.
    
    The probability density of the \rev{target} data $y$\rev{ (the radius fractions of each planetary layer)}, conditioned on the input $x$\rev{ (mass and radius, or mass, radius, and Love number $k_2$ of the planet)}, is represented by a linear combination of Gaussian functions $\phi_i\left(y\;\middle|\;\mu_i(x), \sigma^2_i(x)\right)$:
    \begin{align}
        p\left(y\,\middle|\,x\right) &= \sum^{m}_{i=1}\alpha_i(x)\phi_i\left(y\,\middle|\,\mu_i(x), \sigma^2_i(x)\right),\\
        1 &= \sum^{m}_{i=1}\alpha_i(x),
    \end{align}
    where $m$ is the number of mixture components
    (i.e., the number of Gaussians), 
    $\alpha_i$ are the normalized mixture weights, and $\mu_i$ and $\sigma_i^2$ are the means and variances of the individual Gaussian distributions. \rev{As such, the output layer of the MDN has $3\times m$ neurons: $m$ neurons respectively for the means, variances, and weights of the Gaussian mixture, in each case with their specific activation function.} 
    \revdel{In this study, the features $x$ are the mass and radius of the planets, or the mass, radius, and Love number $k_2$. 
    The target values $y$ correspond to the radius fractions of each planetary layer.}
    
\begin{figure*}[htb]
        \centering
        \includegraphics[width=\textwidth]{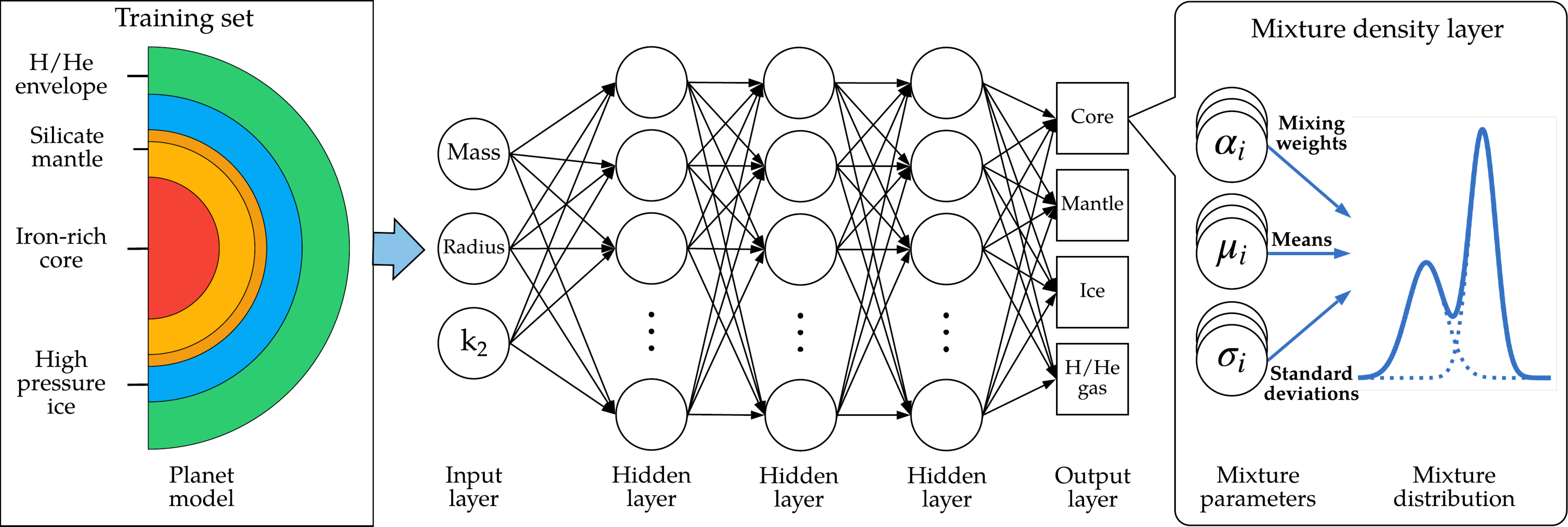}
        \caption{Schematic of the mixture density network architecture used in this work. A large training set of synthetic planets is used to train \rev{a} MDN with 3 hidden layers of 512 nodes each. For each interior layer, the MDN outputs the parameters of a linear combination of Gaussians (mean, standard deviation, and weight of each Gaussian) to fit the distribution of planet interior structures constrained by the input parameters (mass, radius, fluid Love number $k_2$).}
        \label{fig:neural_network}
    \end{figure*}

    Our MDN is built from a feedforward neural network with 3 hidden layers having 512 nodes each.\revdel{We found this setup to be among the best performing MDN architectures by carrying out a grid search of the hyperparameters of the MDN model, varying the amount of hidden layers, the amount of neurons per layer, as well as the learning rate of the optimizer.} 
    For the output layer, we employ a multitask learning approach \citep{caruana1997MultitaskLearning}, where the output layer consists of four seperate MDN layers sharing the same hidden layer (see \cref{fig:neural_network}). The MDN layers\rev{, corresponding to the radius fractions of the four (planetary) layers to be predicted,} are built as specified above using \rev{$m=20$} mixture components each\revdel{, corresponding to the radius fractions of the four layers to be predicted}. \rev{We found this setup to be among the best performing MDN architectures by carrying out a grid search of the hyperparameters of the MDN model, i.e., the amount of hidden layers, the amount of neurons per layer, as well as the learning rate of the optimizer (see below).}
    
    \rev{The hidden layers are activated with a Rectified Linear Unit (ReLU) function \citep{nair2010RectifiedLinear, goodfellow2017DeepLearning}:
    \begin{equation}
        \text{ReLU}(x) = \max(0,x).
    \end{equation}}
    \rev{The ReLU is a commonly used non-linear activation function. Because of their close similarity to linear functions, they are easy to optimize and retain many of the advantages of linear models \citep{goodfellow2017DeepLearning}.}
    
    The MDN layer uses a softmax activation for the $\alpha_i$ nodes\rev{, which ensures that the outputs of those nodes are between zero and one and add up to one.} The $\mu_i$ and $\sigma_i^2$ output nodes are activated with a non-negative exponential linear unit (NNELU) to ensure that predicted means and variances are always positive\rev{ \citep{brando2017MixtureDensity}}. The NNELU is a regular ELU \citep{clevert2015FastAccurate} with an additional offset of +1 to keep it positive: 
    \rev{
    \begin{equation}
        \text{NNELU}(x) = 1 + \text{ELU}(x) = 1 + 
        \begin{cases} 
            \: x & x\geq 0 \\
            \: \exp(x)-1 & x < 0 \quad .
       \end{cases}
    \end{equation}
    }
    
    \revdel{We include a dropout layer after each ReLU layer with a dropout rate of 0.5 \citep{srivastava2014DropoutSimple}. At each training iteration, a dropout layer randomly disables a number of neurons specified by the dropout rate (in our case half of the neurons) in the subsequent layer. Dropout is a common technique to prevent a neural network from overfitting the training data, and thus increases the general robustness of the model.}
    
    \revdel{The hidden layers are activated with a Rectified Linear Unit (ReLU) function \citep{nair2010RectifiedLinear}. The MDN layer uses a softmax activation for the $\alpha_i$ nodes in order to ensure that all weights add up to one. The $\mu_i$ and $\sigma_i^2$ output nodes are activated with a non-negative exponential linear unit (NNELU) to ensure that predicted means and variances are always positive. The NNELU is a regular ELU \citep{clevert2015FastAccurate} with an additional offset of +1 to keep it positive.}
    
    \rev{We choose the loss function of every output layer to be the} average negative log-likelihood  
    \begin{equation}
         \mathcal{L}(\mathbf{\Theta}) = - \frac{1}{N}\sum_{(x,y)\,\in\, \mathcal{D}}\log p\left(y\,\middle|\,x\right),
        \label{eq:nll}
    \end{equation}
    between predictions and input values, where $N$ is the size of the training dataset \rev{$\mathcal{D}$} and \revdel{with} $\mathbf{\Theta}$ \rev{represents} the model parameters. \rev{The model is trained by minimizing the sum of the four individual output losses. }Minimization \rev{of the loss function} is done using the Adam optimizer \citep{kingma2014AdamMethod} with a learning rate of 0.001. \rev{The learning rate defines how much the weights are updated when performing the gradient descent.}
    
    \rev{We include a dropout layer after each ReLU layer with a dropout rate of 0.5 \citep{srivastava2014DropoutSimple}. At each training iteration, a dropout layer randomly disables a number of neurons specified by the dropout rate (in our case half of the neurons) in the subsequent layer. Dropout is a common technique to prevent a neural network from overfitting the training data, because the network will be less sensitive to the weights of individual neurons. This forces neurons to be able to perform well regardless of other neurons in the network, which increases the network's ability to generalize and improves the general robustness of the model.}
    To further help against overfitting, we apply early stopping \citep{prechelt2012EarlyStopping} and stop training once the loss from the validation data does not improve for 15 consecutive training epochs.
    
    \revdel{We test two input scenarios: mass and radius only (case~1); mass, radius, and $k_2$ (case~2).}
    
    The MDN is trained on a NVIDIA Quadro P5000 graphics card using the Keras library \citep{chollet2015Kerasa}, running on top of the TensorFlow library \citep{abadi2015TensorFlowLargeScale}. The code for the MDN layers is adapted from \citet{martin2019CpmpercussionKerasmdnlayer}.\revdel{IPython notebooks to run and use the MDN models are available on GitHub.}
    
    \subsection{Training set}
    We employ our interior structure code to construct each planet model of our training dataset.
    We use a Monte Carlo sampling approach to create a dataset of about $N=900\,000$ unique planet models with random interior structures.
    For each synthetic planet, the planet mass is set to a random value between 0.01 and 25 Earth masses on a logarithmic scale. The mass fractions of the core, ice, and gas layer are also set to a random value, with the mantle layer taking up the remaining mass. \rev{We sample core and ice layers linearly, while the gas mass fraction is sampled logarithmically between $10^{-5}$ and $1$.} A new random draw is performed if the mass fractions add up to more than 1. 
    \rev{In Section \ref{sec:prior} we discuss our choice of sampling distribution.}
    From planet mass and mass fractions, the planet interior structure \rev{and planet radius} \rev{are} calculated. In addition, we calculate the fluid \rev{L}ove number $k_2$ for each planet following the matrix propagator approach of \citet{padovan2018MatrixpropagatorApproach}. 
    We train the neural network with 70\% of the dataset and validate the model results with the remaining 30\%. \rev{We test two input scenarios: mass and radius only (case~1); mass, radius, and $k_2$ (case~2).}
    
    \newpage
    \section{Results}
    
    \subsection{Validation with baseline models}
    
\begin{figure}[htb!]
        \centering
        \includegraphics[width=0.8\linewidth]{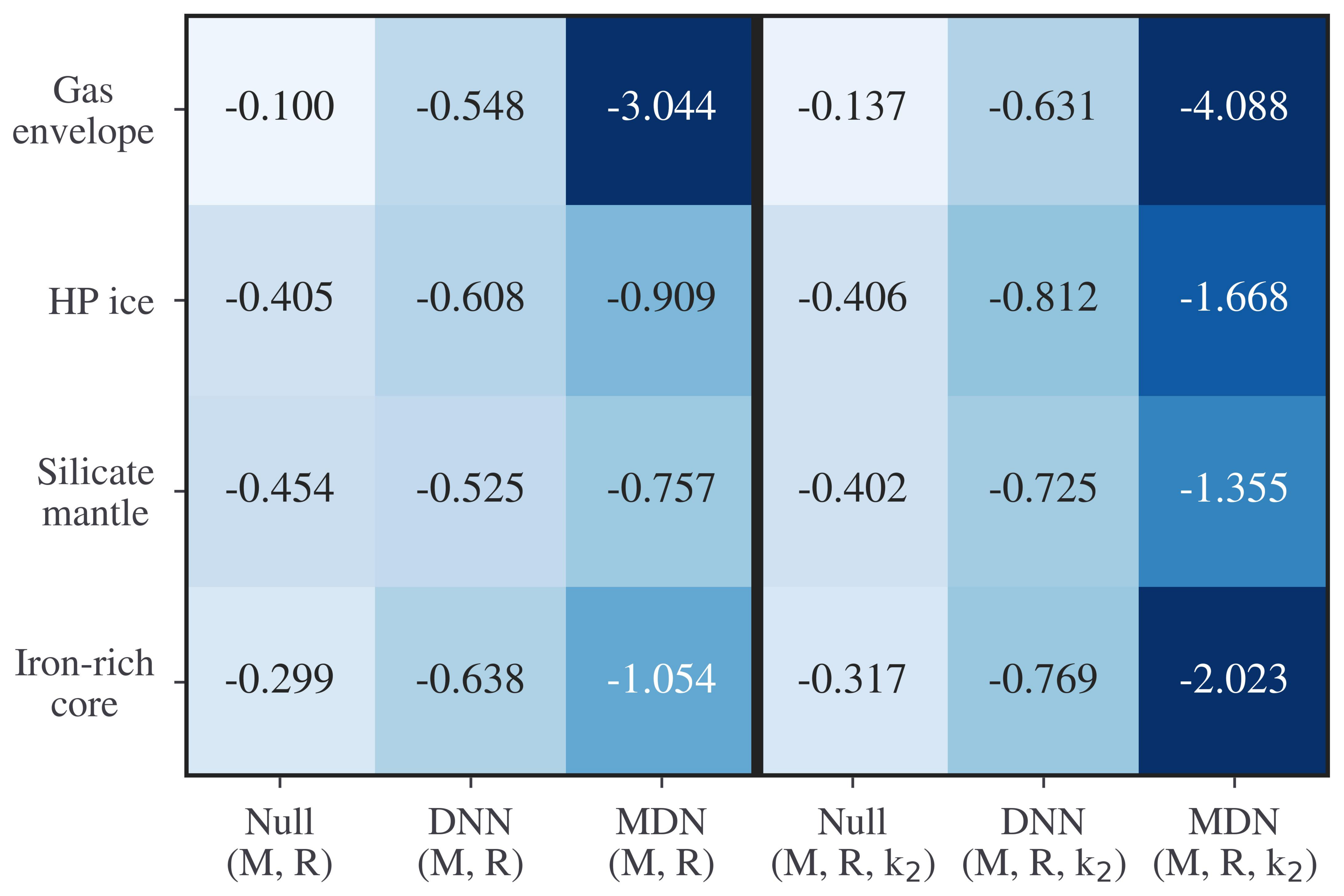}
        \caption{Comparison of the negative log-likelihood $\mathcal{L}(\mathbf{\Theta})$ (NLL) of the MDN and validation models for mass and radius inputs only (left panel) and mass, radius, and $k_2$ inputs (right panel). Lower values \rev{(darker shades of blue)} correspond to a better approximation of the validation data by the model. \revdel{The background colors represent the NLL values for better visibility of differences.}}
        \label{fig:nll_comparison}
    \end{figure}

    We establish a baseline model to evaluate the training performance of the MDN. We use a deep neural network (DNN) with the same parameters and training routine as the MDN, with the exception that the output layer consists of four nodes relating to the radius fraction of the four planet layers. This layer is additionally activated with a softmax activation function \citep{goodfellow2017DeepLearning} to ensure that all output values add up to one, introducing an additional constraint. \revdel{The}\rev{This} DNN is in effect \rev{a} MDN with only one mixture component.
    We also define a null-model that always predicts the mean and standard deviation of layer thicknesses in the training data, thus representing the average interior structure in the data set. The null-model defines the minimum requirements a model must fulfill, in that the thickness of each planet layer must be predicted better than the global average\rev{ of the training data}. A model performing equal or worse than the null-model is basically guessing random values.
    For an indicator of the goodness of the training of the model, we compare the negative log-likelihood (NLL, \cref{eq:nll}) of all three models as shown in
    \cref{fig:nll_comparison}. The three models on the left were trained on mass and radius only (case~1), the three models on the right were trained with mass, radius, and $k_2$ (case~2). Lower values (corresponding to darker \revdel{tones}\rev{shades} of blue) indicate a better performance of the respective models.
    
    \begin{figure}[htb!]
        \centering
        \includegraphics[width=0.8\linewidth]{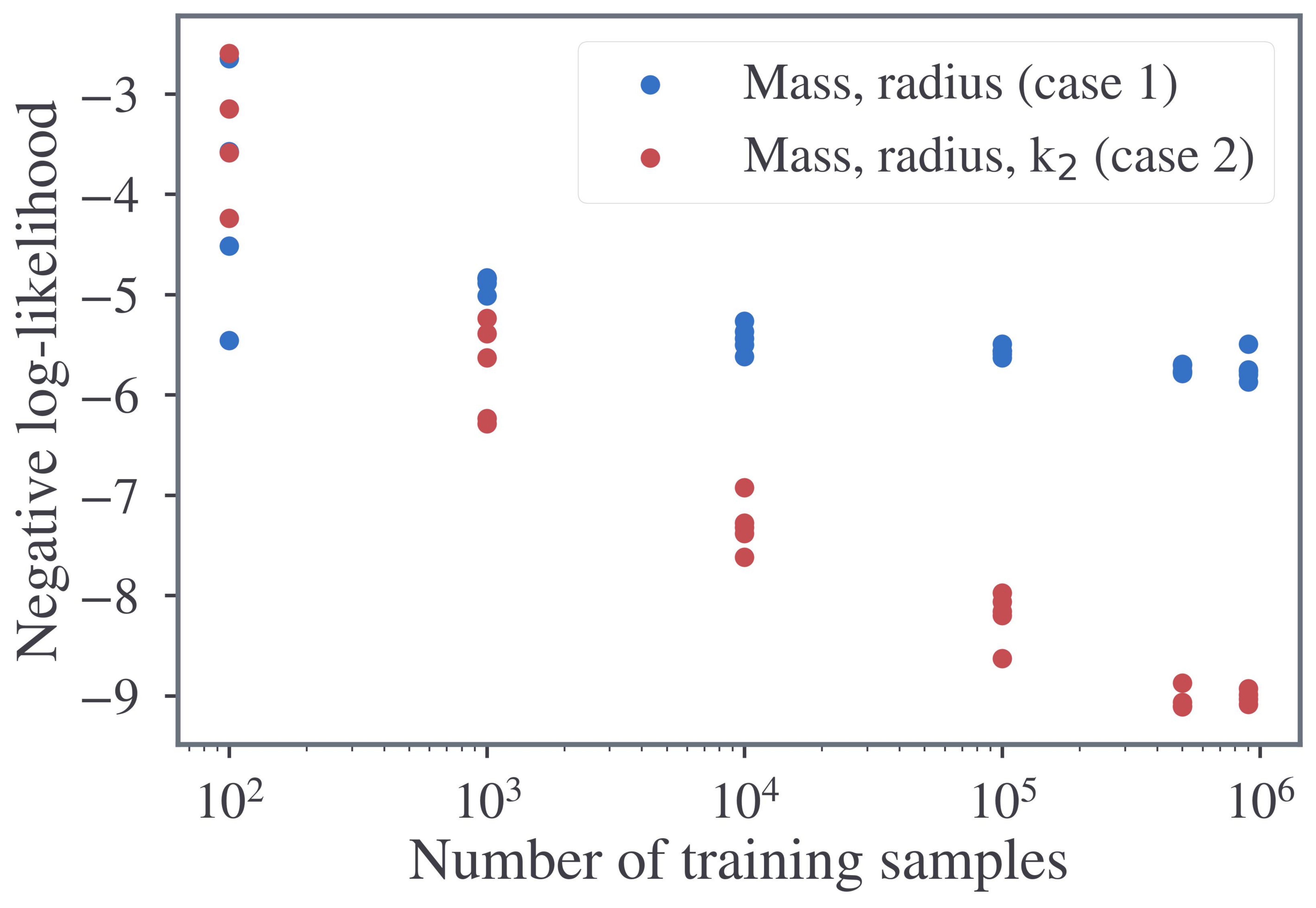}
        \caption{Dependence of the \revdel{average} NLL on the size of the training set for the MDN using only mass and radius inputs (blue dots), and for the MDN using mass, radius, and $k_2$ (red dots).\rev{ For each training set size, we changed the random seed to generate a different set of training samples.}}
        \label{fig:nll_convergence}
    \end{figure}
    
    Both the DNN and MDN perform better than the null-model for all planetary interior layers. The MDN shows significant improvement for all interior layer predictions compared to the DNN due to its ability to accommodate multiple interior solutions for one mass and radius input. 
    
    In \cref{fig:nll_convergence} we demonstrate the dependence \rev{of the NLL} \revdel{training} loss on the size of the training data set.\rev{ For each training size, we changed the random seed to sample a different set of training data.} For both case~1 and case~2, the loss improves with increasing sample numbers, \rev{and the spread of loss values is reduced.}\revdel{although the} \rev{The NLL flattens} for higher sample numbers. We conclude that a training set size of $N\approx10^5$ is sufficient to ensure a good fit of the MDN. Case~2, using mass, radius, and $k_2$, needs more training samples than case~1, due to the increased number of observables.
    
    \subsection{MDN accuracy and degeneracies}
    
    \begin{figure*}[htb!]
        \centering
        \includegraphics[width=\linewidth]{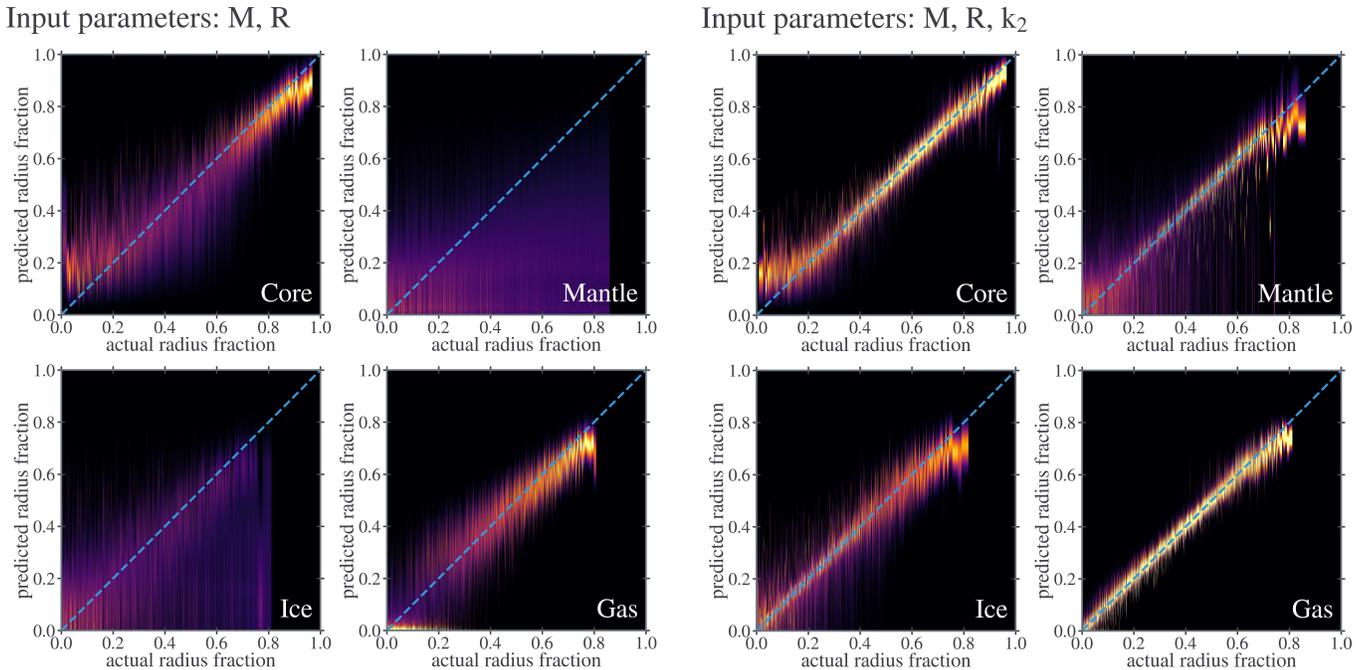}
        \caption{Predicted distributions of the thickness of core, mantle, ice, and gas envelope layers as a function of actual values from the validation data set. The dashed diagonal line indicates a perfect prediction of the layer thickness. The four panels on the left show predictions for \rev{a} MDN trained on mass and radius inputs only. The four panels on the right show the predicted distributions for \rev{a} MDN trained additionally with the fluid Love number $k_2$. The predicted distributions are coloured according to the local probability density.}
        \label{fig:confusion}
    \end{figure*}
    The ability of the network to constrain the interior structure of each planet based on given input parameters is displayed in  \cref{fig:confusion}, which shows the distributions predicted by the MDN from the validation data set of each interior layer plotted against the actual relative thickness of the layer. \rev{A} MDN model that perfectly predicts the actual data set would plot along the diagonal.
    
    With only mass and radius as inputs, the MDN can infer the relative thickness of the gaseous envelope well. This result is similar to the findings of \citet{lopez2014UnderstandingMassRadius}, who showed that the planetary radius can be used as a first-order proxy for the mass of the H/He envelope. Likewise, the iron-rich core is relatively well constrained. For model planets with large cores, the predictions are quite accurate, as the high density of these planets does not allow for much variation in possible interior structures, given that we do not allow for planets denser than a naked core. The same applies to planets with thick gas envelopes, as there is nothing less dense than the gas. The MDN tends to overestimate the core thickness when the core is small. The reason for this behaviour lies in the nature of the mass-radius diagram: For planets with small cores, a slight increase of the core mass leads to a larger relative increase in core size than for planets with massive cores, leading to a higher degeneracy of the interior structures. \rev{As such, very small core sizes are undersampled in the training data set (see also \cref{fig:cdf}). This makes the core size of small core planets more difficult for the MDN to predict. Similarly, the thickness of large layers tends to be underestimated due to less training samples available.}
    
    The high-pressure ice layer thicknesses are constrained relatively well for the most part, but the MDN tends to underestimate the thickness of the ice layer in some parts of the dataset. The silicate mantle is barely constrained. The MDN mostly predicts a broad distribution with constant mean values, but performs slightly better for planets with small mantles. This is not an unexpected result, as it is a consequence of the inherent degeneracy in the mantle and ice layer thickness due to \rev{both layers having intermediate densities in comparison with other layers in the planet. As such, their masses and radii can easily be compensated by those other layers}. \cref{fig:confusion} shows that the thickness of mantle and ice layers is barely constrainable for most planets.
    
    The inclusion of $k_2$ dramatically improves the performance of the MDN, which now adequately predicts the correct thickness of each interior layer (right panels of \cref{fig:confusion}). The biggest improvement is in the prediction of the mantle layer.
    However, the mantle remains the least well predicted  layer as already shown by the scores of \Cref{fig:nll_comparison}. The MDN still tends to overestimate the thickness of the core, mantle and ice layer when these layers are thin \rev{due to undersampling in the training data set}. The previous underestimation of the ice layer is no longer observed. The gas layer thickness is predicted with a high accuracy, even for thin layers. All in all, the inclusion of $k_2$, by providing a measure of the degree of mass concentration in the interior, significantly constrains the solution space and helps reduce the degeneracy of interior solutions.
    
    \subsection{Application to Solar System and extrasolar planets}
    
    \begin{figure*}[htb!]
        \centering
        \includegraphics[width=\linewidth]{MR_predictions.pdf}
        \caption{MDN predictions of the relative interior layer thickness for four representative planet cases using only mass and radius as inputs to the network. The colored lines show the combined Gaussian mixture prediction of the MDN. \revdel{Each Gaussian mixture is normalized so that the area under the curve adds up to one}\rev{The area under each curve is normalized to one}. The histograms show possible interior solutions within 5\% of the input values obtained from\rev{ an} independent Monte Carlo sampling\rev{ (see main text for more details)}. The colored bars in the upper plots represent the space of solutions of valid interior structures obtained from \rev{the same }Monte Carlo sampling. The shade of each bar corresponds to \revdel{the allowed error}\rev{an uncertainty} of 1\%, 2\%, and 5\% in the observable parameters (from darkest to lightest, respectively). The vertical dashed lines in the Earth plot show the actual thickness of Earth's core and mantle from PREM \citep{dziewonski1981PreliminaryReference}. Data for Kepler-10b from \citet{weiss2016REVISEDMASSES}. Data for GJ 1214b from \citet{harpsoe2013TransitingSystem}.}
        \label{fig:MRpredictions}
    \end{figure*}
    
    \begin{figure*}[htb!]
        \centering
        \includegraphics[width=\linewidth]{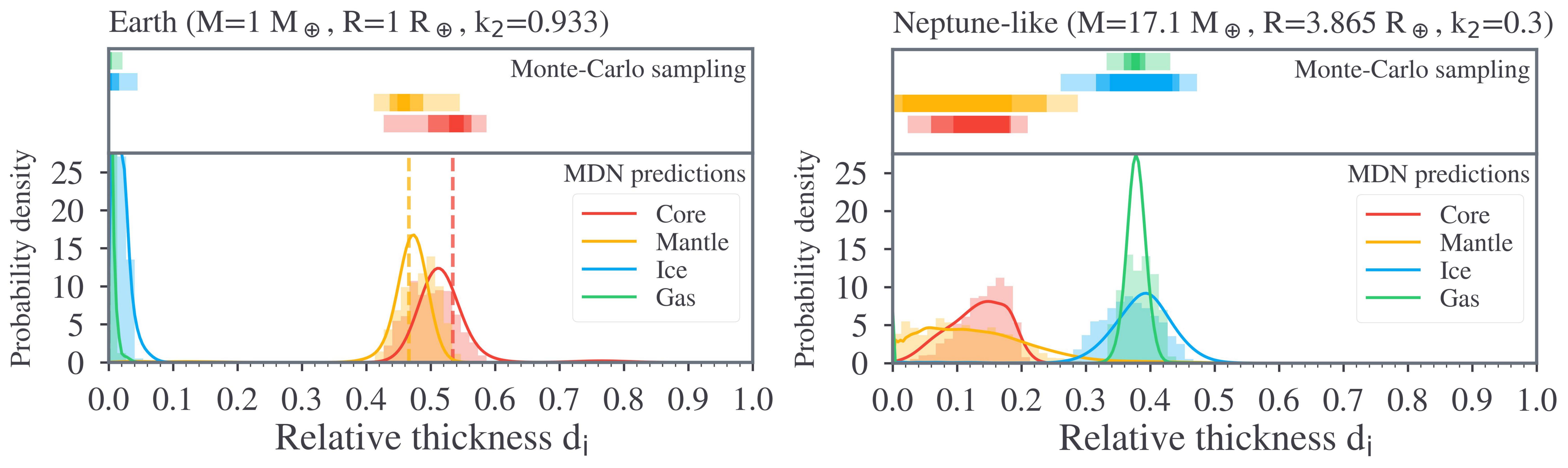}
        \caption{MDN predictions of the relative interior layer thickness for the Earth (left) and \rev{a Neptune-like planet with $k_2=0.3$} (right) \rev{as shown in} Figure \ref{fig:MRpredictions}, but using mass, radius, and the fluid Love number $k_2$ as inputs to the network.} 
        \label{fig:MRk2predictions}
    \end{figure*}
    
    We investigate the predicted interior structures for four representative planets. In the Solar System, we consider the Earth, being the archetypical terrestrial planet, and Neptune as a gas and ice-rich planet at the upper end of our mass range (with a mass of $17.1$~M$_\oplus$ and a radius of $3.865$~R$_\oplus$). For exoplanets, we investigate Kepler-10 b (3.72~M$\oplus$, 1.47~R$_\oplus$, \citealp{weiss2016REVISEDMASSES}), which with its high density is a \revdel{potential rocky}\rev{likely} Super-Earth \citep{valencia2010CompositionTransiting, batalha2011KEPLERFirst, wagner2012RockySuperEarth}, and GJ 1214 b \citep{charbonneau2009SuperEarthTransiting, rogers2010ThreePossible, nettelmann2011ThermalEvolution}, a well-studied, relatively low-density planet (6.26~M$\oplus$, 2.85~R$_\oplus$, \citealp{harpsoe2013TransitingSystem}), likely with a volatile-rich interior and extended gas envelope.

    
    \Cref{fig:MRpredictions} shows the distributions of predicted solutions of the MDN for each interior layer for the four planets using mass and radius only as inputs for the network. The histograms behind each prediction curves show all interior solutions within a  5\% margin of mass and radius\rev{, assuming a uniform error distribution where every input parameter is equally likely.} \rev{To obtain these samples, we modeled an independent data set by Monte Carlo sampling from the same prior distribution of mass fractions that was used to create the training data set.} Additionally, the bars on top of each plot show the \revdel{range}\rev{spread} of possible solutions obtained from this Monte Carlo sampling, where the color shade \revdel{refers to}\rev{shows all samples within the} 1\%, 2\%, and 5\% error margins in input parameters (from darkest to lightest). The \rev{relative error $\delta$ of each Monte Carlo sample is} \revdel{are} calculated as the \revdel{relative} Euclidean distance from the input parameters\rev{:
    \begin{equation}
        \delta = \sqrt{\left(\frac{\Delta M}{M}\right)^2 + \left(\frac{\Delta R}{R}\right)^2},
    \end{equation}
    where $M$ and $R$ are the true mass and radius values, and $\Delta M$ and $\Delta R$ the absolute errors between the Monte Carlo sample and true value.}
    The comparison of the MDN predictions (solid lines) and the Monte Carlo distributions (histograms) in \cref{fig:MRpredictions} show that the MDN accurately replicates the distributions of possible interior solutions for all planets.
    
    In the case of the Kepler-10b, the predicted core size is consistent with values from other studies \citep[e.g.,][]{wagner2012RockySuperEarth}, showing it to be most likely a terrestrial type planet. In contrast, the predicted distributions of GJ 1214b are more similar to Neptune's interior. This compares well with the study of \citet{nettelmann2011ThermalEvolution}, who conclude that the interior of
    GJ 1214b is consistent with having a large gaseous, water-enriched envelope. Other possible interior structures \rev{in their study} include the planet being mostly rocky with a large H/He envelope taking up 40\% of the planet's radius.
    
    For the Earth, the dashed lines in \cref{fig:MRpredictions} indicate the actual core and mantle thickness as known from the Preliminary Reference Earth Model \citep{dziewonski1981PreliminaryReference}. These do not align with the peaks of the predicted MDN distributions, which reproduce the correct values only near their tails (\Cref{fig:MRpredictions}). This is a result of the large degeneracy in interior solutions. A planet like the Earth, consisting of iron core and silicate mantle only, is only one solution (in terms of possible solutions of the interior structure) compared to a generic planet having ice and gas layers. The situation would partly improve with additional constraints on the planet, such as transmission spectra or detailed escape modeling ruling out the presence of a substantial atmosphere\rev{ \citeg{owen2017EvaporationValley, dorn2018SecondaryAtmospheres, kubyshkina2018GridUpper}}.
    
    \subsection{Influence of $k_2$ on the predictability of the interior structure}
    
    As shown in \cref{fig:confusion}, adding the fluid Love number $k_2$ as a third observable helps to significantly constrain the interior structure of the planets.
    \cref{fig:MRk2predictions} (left) shows the predictions of the neural network including the fluid Love number $k_2$ for the Earth (with $k_2=0.933$ from \citealt{lambeck1980EarthVariable}). For Neptune (right), the actual value of $k_2$ is $0.392$ calculated from the first order of the expansion of the gravitational moment $J_2=3.40843\e{-3}$ \citep{jacobson2009ORBITSNEPTUNIAN}, using the volumetric mean radius of $24.622\e{6}$~m and a Voyager rotation period of $16.11$~h \citep{hubbard1984PlanetaryInteriors}. \revdel{In this work, however, we use a value of $k_2=0.3$, as the assumption of the zero-temperature EoS puts the real mass, radius, and $k_2$ triplet of Neptune outside of our sampled parameter space, preventing us from reaching the actual value of $k_2$. A more general, temperature-dependent EoS of the gas envelope could allow us to model Neptune more closely to the correct $k_2$.}\rev{In this work, however, the assumption of the zero-temperature EoS for the gaseous envelope puts the real mass, radius, and $k_2$ triplet of Neptune outside of our sampled parameter space, preventing us from reaching the actual value of $k_2$. As mentioned in Section \ref{sec:envelope}, our model overestimates the thickness of the gas envelope when constraining the planet's interior to mass and radius. As a result, planets with significant gas fractions are more centrally condensed in our model, resulting in low values of $k_2$. For a planet with Neptune mass of 17.1~M$_\oplus$ and radius of 3.865~R$_\oplus$, possible values of $k_2$ lie between 0.2 and 0.325 with our model. As such, we chose to use a value of $k_2=0.3$, which allows the interior model to return a good spread of solutions while still being close to the actual value. We call this planet \textit{Neptune-like}, noting that a more sophisticated, temperature-dependent EoS of the gas envelope would be needed to model Neptune in better detail and with the correct $k_2$.}
    \revdel{The value of $k_2=0.3$\rev{This value} is chosen such \rev{as} to allow the interior model to return a good spread of solution\rev{s} while still being close to the actual value.}
    
    In both cases, the inclusion of $k_2$ significantly shrinks the solution space.
    For the Earth, the interior structure of the Earth is constrained well to the actual values. In the case of \rev{the Neptune-like planet}, the MDN predictions clearly indicate \rev{this planet} to be an ice-rich planet with an extended gas envelope and a small iron-silicate core. The predictions are also consistent with \rev{the planet} having no iron-silicate core at all. The gas-envelope thickness\revdel{ of Neptune}, \rev{assuming} a Love number of 0.3, can be well constrained by the MDN.
    
    Compared to the 3-layer interior structures models of Neptune by \citet{podolak1995ComparativeModels} and more recent ones by \citet{nettelmann2013NewIndication}, which show that Neptune's ice-enriched gaseous envelope makes up 10--30\% of the planet's radius, the MDN overestimates the thickness of the gas envelope to be 30--60\% of the planet's radius when considering mass and radius only (\cref{fig:MRpredictions}), and 30--44\% of the planet's radius when considering mass, radius, and $k_2$ (\cref{fig:MRk2predictions}). There are several reasons for this finding. For one, we use a basic H/He atmosphere with no temperature-dependence of the envelope, and a pure water ice layer underneath. Compared to interior density profiles of \citet{nettelmann2013NewIndication}, we overestimate the density of the ice layer by $\approx 25\%$ in the lower parts of the planet. This is a result of our choice of a high-pressure EoS neglecting ionic and super-ionic water phases. As a result the gas envelope needs to be larger to arrive at the same mass and radius, compensating the denser interior (see section \ref{sec:envelope}). Compared to more recent studies, the zero-temperature EoS that we use overestimates the density \rev{especially in} the upper part of the atmosphere. The combination of these two factors causes the envelope to be comparatively larger than currently believed. In turn, this also leads to our planet models being more centrally condensed, so that $k_2$ tends to be underestimated in all models with extended atmospheres.\revdel{ Nevertheless, for exoplanets, the metallicity of the atmosphere is mostly unknown.}
    
    Additionally, our interior model assumes strictly immiscible layers. Some layer phases, such as H$_2$ and H$_2$O, can actually be miscible under pressure and temperature conditions in gas giants. As such, the layered interior model can not fully account for the interior of Neptune \citep[see also][]{podolak1995ComparativeModels}.
    
    The Neptune\rev{-like} predictions of the MDN are however fully consistent with the Monte Carlo sampling results. As such, deviations from the actual interior structure are only due to simplifications in the \rev{interior} model, \rev{reflected} in the training data. Therefore, we expect that using a more appropriate atmosphere and an improved ice model will help \rev{us to} improve the accuracy of our predictions.
    
\begin{figure*}[htb!]
        \centering
        \includegraphics[width=\linewidth]{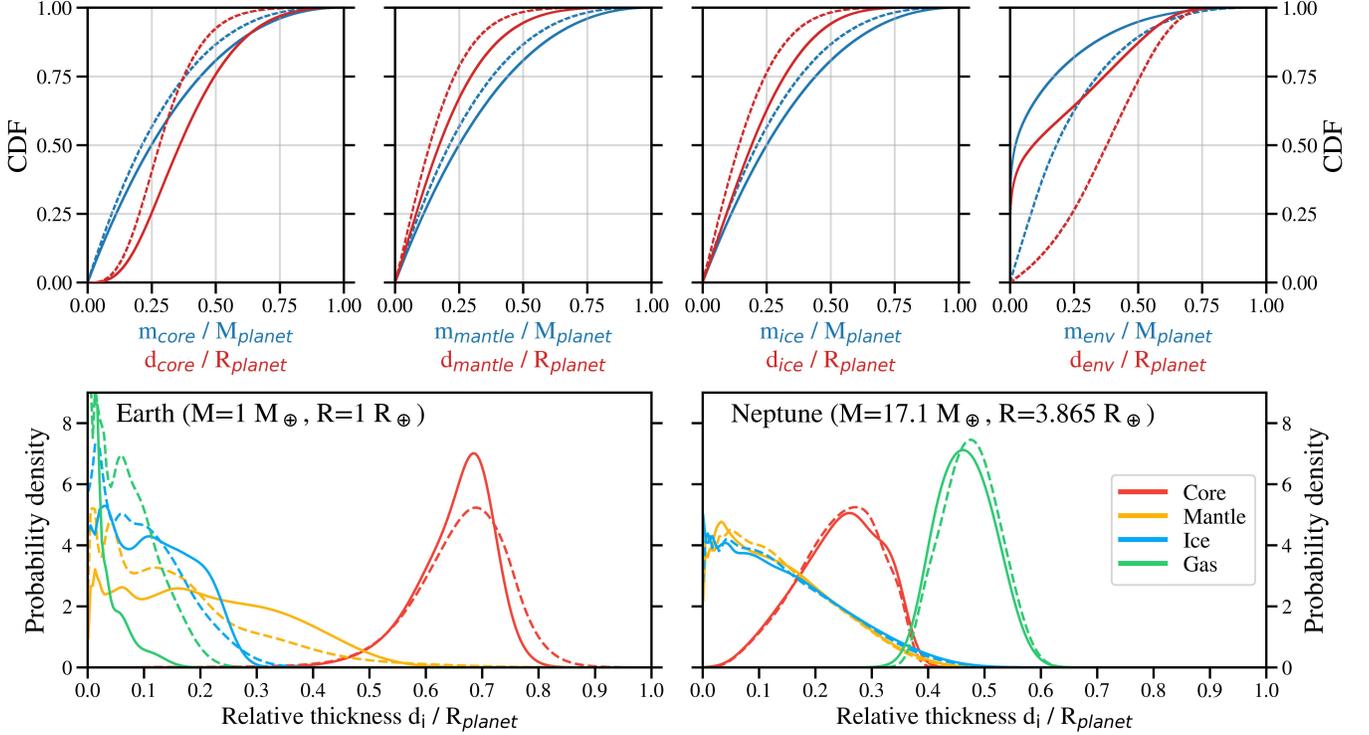}
        \caption{\rev{Top plots: Cumulative density function (CDF) of the training data set for layer mass fractions (blue) and layer radius fractions (red). Solid lines represent the distribution of the full training set with a logarithmic sampling of the envelope mass fractions. For the dashed lines, the dataset was resampled so that the distribution of gas mass fractions approximates the distribution of the other layers' mass fractions (with linear sampling). Bottom plots: Comparison of predictions for both prior distributions for Earth and Neptune (using mass and radius as network inputs).}}
        \label{fig:cdf}
    \end{figure*}

    \rev{
    \subsection{Influence of prior distribution}\label{sec:prior}
    Since the MDN predicts the posterior distribution of possible planetary structures solely from the data it was given during training, the distribution of training data points can be a main source of prediction errors. In \cref{fig:cdf}, we show the effect of a different prior choice on the predicted interior structures of Earth and Neptune. Solid lines represent the baseline prior distributions in mass (blue) and radius (red), where mass sampling is linear except than for the gaseous envelope, where it is logarithmic. The dashed lines are the priors obtained assuming a linear mass sampling in each layer. This drastically reduces the amount of planets with a small envelope (red lines), shifting focus to planets with more extended envelopes. This slightly changes the priors of the other layers as well.\newline
    The bottom plots in \cref{fig:cdf} show a comparison of the predictions of a MDN trained with this new resampled data set with the predictions of the original MDN. For the Earth case, the greatest difference lies in the prediction of the envelope thickness. Here, the MDN tends to predict a preferentially thicker envelope. This is due to the new prior giving less priority to very small envelopes. As such, there are less training samples available for small envelopes. As a result of the changed envelope prediction, the predictions of the other layers also change slightly. Mantle and ice predictions shift to slightly lower thicknesses, the core shifts to slightly larger thicknesses. However, the predicted interior structures still lie in a similar range, only the shape of the curves are different. For Neptune, there is very little difference between predictions for the new and old priors, since planets with larger envelope fractions are equally well sampled with both prior distributions. We conclude that the exact shape of the prior distributions is not too important as long as there are enough samples in each region of the solution space.}
    
    \newpage
    \section{Discussion and conclusions}
    
   We show that mixture density networks (MDNs, Figure \ref{fig:neural_network}) can accurately 
   predict
   the  space of interior solutions of planets 
   with mass below 25 Earth masses
   (Figure \ref{fig:MRpredictions}). There are two main advantages in 
   this machine-learning-based approach
   compared to conventional methods
   (e.g., Monte Carlo sampling): 
   First, predicting the distribution of possible interior structures of a planet with a trained MDN only takes few milliseconds on a normal desktop CPU (we used an AMD EPYC 7351) compared to potentially several hours if the parameter space needs to be sampled with a forward model.
   Thus, 
   trained MDNs can be used to provide a rapid first characterization of the space of admissible interior structures of observed planets. These can be then investigated with more sophisticated models that account for 
   additional observables relevant to
   specific planets such as stellar irradiation, planet age, and atmospheric composition. Secondly, the trained MDN 
   represents a stand-alone tool:
   While its predictions rely on the assumptions of the underlying interior model, the interior model itself as well as the training data set are not needed to operate the MDN. 
   This renders MDNs a valuable tool for the exoplanetary science community because a fully trained MDN can be used to obtain an indication of possible interior structures without the need of a dedicated code for solving the forward problem. The file size of our MDN is $\approx 7.5$ MB, making its dissemination easy. \rev{Our trained MDN models are available on GitHub\footnote{\url{https://github.com/philippbaumeister/MDN_exoplanets}}.}
    
    The generation of a data set of forward models of appropriate size can be computationally expensive, especially if a complex interior model is used or if additional observables are considered (Figure \ref{fig:nll_convergence}). Conversely, the training of the network itself only takes a few minutes, and can be significantly sped up by utilizing GPU acceleration \citep{shi2016BenchmarkingStateoftheArt}. Therefore, 
    several MDNs corresponding to
    different combinations of input parameters 
    (i.e., observables)
    can be efficiently tested, making it possible to quickly identify combinations of parameters that better characterize the interior of a planet.

    \rev{Similar to other inference methods, the} predicted distributions depend to some extent on the choice of the prior distributions \citep[see, e.g.,][]{dorn2017GeneralizedBayesian}.
    \rev{With better informed priors constrained by observations and physical considerations, the predictions will be more accurate.}
    \revdel{When more informed priors can be considered 
    \citep{Kreidberg2019} 
    the predictions will be more accurate.}
    The MDN approach presented here reproduces the well-known
    degeneracy of interior structures based
    on mass-radius pairs only (Figures \ref{fig:confusion} and \ref{fig:MRpredictions}).
    Using $k_2$ as an additional input significantly reduces the set of possible interior structures (Figures \ref{fig:confusion} and \ref{fig:MRk2predictions}).

    \revdel{However, the same assumption applied without $k_2$ has been 
    show\rev{n} to provide poor results, since 
   the formation history as well as stochastic processes like giant impacts can alter the planetary mineral composition 
    \citep{dorn2015CanWe}.}

    In the solar system there are only  
    few planetary objects, 
    each characterized by 
    several remote and in-situ observations that 
    provide a relatively accurate picture of their 
    formation, history, and interior structure.
    The problem of understanding exoplanets is 
    complementary, in that thousands of objects are known,
    each characterized by a severely limited number of
    observations.
    MDNs provide \rev{a fast alternative to other, often time-consuming, inference methods, allowing an efficient approach to exploiting exoplanetary data, which span a parameter space much wider than that of the planets of the solar system.} 
    As new type of data will become available with
    the advent of new observing facilities (e.g.,
    atmospheric compositions from JWST), models more
    complex that those presented here, and larger data
    sets of forward models to
    train appropriate MDNs will be required.
    These are modifications that only depend on the
    available computing power, and can be readily
    incorporated in the method presented here
    as the need arises.
    
\acknowledgements
    We thank an anonymous reviewer for insightful comments that helped improve the paper.
    We describe contributions to the paper using the taxonomy of \citet{brand2015}.
    For each entry, the first name is lead, the following are contributing.
    {\it Conceptualization:} S.P., N.T.; {\it Methodology:} P.B., S.P., N.T., G.M., M.G.; {\it Software:} P.B., S.P.; {\it Investigation:} P.B.; {\it Writing--Original Draft:} P.B., S.P., N.T.; {\it Writing--Review \& Editing:} P.B., S.P., N.T., G.M., N.N., J.M., M.G.; {\it Visualization:} P.B.; {\it Supervision:} N.T., S.P., M.G.; {\it Funding Acquisition:} N.T., M.G.\newline
    The authors acknowledge the support of the DFG Priority Program SPP 1992 ``Exploring the Diversity of Extrasolar Planets'' (TO 704/3-1, GO 2610/2-1) and the DFG Research Unit FOR 2440 ``Matter under planetary interior conditions''. N.T. also acknowledges support by the Helmholtz Association (VH-NG-1017), and M.G. by the DFG (GO 2610/1-1).

\bibliography{references,Additional_Ref}

\listofchanges
\end{document}